# Activity of Pd$_n$ (n=1-5) clusters on alumina film on Ni$_3$Al(111) for CO oxidation: a molecular beam study

Georges Sitja and Claude R. Henry

Aix-Marseille Université, CNRS, CINaM, F-13288 Marseille, France

Single atom catalyst (SAC) is a vivid new area of research in catalysis. However, the activity in CO oxidation of isolated Pt or Pd atoms, generally supported on an oxide powder, is still controversial. Furthermore, the steady state activity of few atoms clusters is still not yet quantitatively known. In this work we study, by molecular beam reactive scattering (MBRS), the activity of Pd$_n$ (n= 1-5) clusters, grown on a nanostructured alumina films on a Ni$_3$Al(111) surface. It is shown that the single atoms are not active at 473 K but they diffuse and coalesce at 533K forming larger clusters. The activity of a cluster is proportional to the number of atoms it contains (n=2-5). At 533 K, the activity per (surface-) atom, which is the turnover frequency (TOF), is constant. Its value is close to those obtained for large clusters of 181±13 atoms and Pd (111) extended surfaces, in the same experimental conditions.

## I. Introduction

A key question in catalysis is: how to reduce the quantity of precious metals in catalysts which are rare and expensive? A tentative answer was given in 2011 by the discovery that single Pt atoms supported on iron oxide powder could catalyze CO oxidation [1]. Indeed this paper showed, for the first time, that isolated Pt atoms identified by HAADF-STEM, before the reaction, were more active near room temperature than gold nanoparticles on Fe$_2$O$_3$ (a reference catalyst for CO oxidation at RT which is much more active than Pt particles catalyst at this temperature). Since the publication of this paper an explosion of publications appears on this new topic called 'Single Atom Catalyst' (SAC). An excellent review of this topic has been published recently by Liu and Corma [2].Other metals than Pt like Pd, Ni, Ru, Rh, Au, Ag …. have shown single atom activity for various reactions but CO oxidation remains the most studied reaction on SACs [2]. Pt$_1$ supported on alumina showed a smaller activity than on iron oxide [3]. Pd$_1$ supported on -alumina was also found active in CO oxidation [4]. However, some controversy appears about the catalytic activity of SACs. An infrared study showed that on a Pt/HZSM-5 catalyst only Pt clusters are active in CO oxidation and water-gas shift whereas that Pt$_1$ is only spectator [5]. More recently, two *in situ* studies during reaction have shown that Pt clusters and nanoparticles supported on Al$_2$O$_3$ and TiO$_2$, are more active in CO oxidation than Pt single atoms [6, 7]. In fact, these studies show that during reaction Pt atoms are not stable and coalesce forming Pt clusters. Other studies, using STEM or infrared spectroscopy, have shown that Pt atoms on TiO$_2$ nanoparticles are stable under oxygen but diffuse under hydrogen at 450°C [8], that Pt atoms on Fe$_2$O$_3$ coalesce under CO or H$_2$ [9] and that Pd atoms supported on -Al$_2$O$_3$ coalesce under CO oxidation [10]. A mean to avoid coalescence of Pt atoms is to isolate Pt atoms on TiO$_2$ nanoparticles (5 nm) [11]. In the last study the turnover frequency (TOF) for CO oxidation on Pt single atoms was found two times larger than on 1 nm particles [11]. Besides these controversies, several problems remain unsolved with SACs. First, the anchoring sites for metal atoms which certainly play a role in their catalytic activity are generally unknown that is a severe limitation for a valuable comparison between experiment and theory. Second, the actual dispersion of the catalyst is not always accurately known (for example a few metal clusters can have a much higher activity than single atoms) that limits an accurate determination of the TOF. Third, there is a lack of knowledge about the evolution of the catalytic activity as a function of the number of atoms forming the cluster. This last point can be in principle solved by experiments with size selected deposited clusters [12-17]. In 1999, U. Heiz and co-workers studied by temperature programmed reaction (TPR) the CO oxidation of Pt$_n$ (n= 5-20) clusters trapped on point defects of the surface on an MgO film [12]. The activity (obtained by integration of the TPR peaks) was found dependent of the number of atoms in the clusters but it was not possible to measure the steady sate activity by this method. The same group tried to measure by TPR the activity in CO oxidation of Pd single atoms deposited on the same MgO surface [13]. However, during the first TPR cycle Pd single atoms moved and coalesced [13]. By using a pulsed CO valve TOF and reaction probability (RP) measurements, for CO oxidation, have been obtained on Pd$_8$ and Pd$_{30}$ clusters on MgO (100) films [14]. Recently, the group of Anderson measured the true steady state activity in CO oxidation of Pd$_n$ (n= 5, 20, 25) clusters supported on alumina films [15]. In

2015 the group of Yasumatsu has studied the CO oxidation on $Pt_n$ (n=10, 30, 60) supported on Si(111) [16]. Later, Watanabe and coworkers have studied by TPR the CO oxidation of $Pt_n$ (n = 7-35) clusters supported on $Al_2O_3$/NiAl (110) ultrathin films [17].

In the recent years surface science studies, using mainly STM and TPR techniques, have been devoted to the reactivity of single atoms deposited on oxide surfaces. Parkinson and coworkers have shown that isolated Pd atoms are stable up to 673 K on $Fe_3O_4$ (100) surface under UHV while under CO they are mobile at RT [18] as well as Pt atoms [19]. The oxidation of methanol has been also studied by TPR on Pd single atoms on $Fe_3O_4$(001) but at RT Pd atoms coalesce in the presence of methanol [20]. In order to stabilize atoms or small clusters on a surface, several groups have tried to use templates having a regular array of defects acting as nucleation centers and anchoring sites for clusters [21]. Indeed, perfects arrays of Sm and Dy atoms have been obtained by atom deposition at 40 K on the moiré structure formed by a graphene monolayer on Ir(111) [22] but deposited atoms become mobile already at 78 K. On the same graphene template, the group of Michely showed that it was not possible to stabilize clusters smaller than 5 atoms at RT and moreover the clusters became mobile under CO or $O_2$ [23]. The same group used another template h-BN/Ir (111) allowing stabilization of Ir clusters containing 6 to 175 atoms up to 700K under UHV [24]. On h-BN/Rh(111) another group has prepared at RT, under UHV, Pt single atoms and $Pt_n$ clusters (n=2-50) [25]. Pt clusters containing around 12 atoms were stable up to 400 K [25]. Another study with the same template has shown that Pd single atoms are mobile and coalesce at RT under UHV [26]. A third template, formed by an ultrathin film (0.5 nm) of alumina obtained by high temperature oxidation of a $Ni_3Al$ (111) surface, has been discovered by Becker et al [27] and used to grow arrays of clusters (Cu, Ag, Au,….). On this template Pd clusters containing 5 to 400 atoms are stable up to 600K under UHV and under CO, at least up to 500K, allowing the measurement of the desorption energy of CO as a function of cluster size [28, 29]. The CO oxidation has been studied on such arrays of Pd clusters (181 atoms) by MBRS (molecular beam reactive scattering) at steady state [30] and under transient conditions to determine the Langmuir-Hinshelwood barrier as a function of cluster size (174, 360 and 768 atoms) [31]. The surface structure of ultrathin alumina film on $Ni_3Al$ (111) has been investigated at atomic resolution by nc-AFM and STM [32, 33]. This alumina film presents two hexagonal superstructures formed by well-defined sites which are nucleation centers for the metal clusters. Pd clusters sit on the 'dot' structure which has a density of sites of $6.5 \times 10^{12}$ $cm^{-2}$ while V clusters are formed on the 'network' structure with a density of sites of $1.95 \times 10^{13}$ $cm^{-2}$ [27]. A third hexagonal superstructure called 'dot/3' and corresponding to a density of $5.85 \times 10^{13}$ clusters per cm², has been recently discovered by growing small Pt clusters (1-6 atoms) at RT, while by growth at 573 K the Pt clusters sit on the 'network' structure [34].

In this paper we form, by atom deposition, very small Pd clusters (1-5 atoms) on the 'dot' structure of $Al_2O_3$/$Ni_3Al$ (111) and we measure the steady state rate of CO oxidation at different temperatures by MBRS. From a careful analysis of the $CO_2$ production as a function of the mean number of atoms per site of the 'dot' structure we show that Pd single atoms are inactive and that the activity of larger clusters is proportional to the number of atoms they contain.

**II. Experimental**

2.1. Sample preparation

$Ni_3Al$ (111) surfaces are prepared by several cycles of Ar ion sputtering (current = 15 A, energy = 2 keV, time = 15 min) followed by annealing at 1100K during 20 min. The sample temperature is accurately measured by a thermocouple in intimate contact with the $Ni_3Al$ (111) single crystal and by an optical pyrometer. The alumina film is obtained by oxidation of the $Ni_3Al$ (111) surface at 1000K under oxygen ($P_{O2}$ = $5 \times 10^{-8}$ mbar, exposure = 45 L). The quality of the alumina film is checked *in situ* by low energy electron diffraction (LEED) [28]. Pd clusters are grown by deposition at 373 K of Pd atoms from a Knudsen cell. The Pd flux ($2.30 \pm 0.03 \times 10^{12}$ atoms/s.$cm^{-2}$) is *in situ* calibrated before each deposition by a quartz crystal microbalance. The quality of the organization and the reproducibility of the arrays of Pd clusters has been checked *in situ* by STM and GISAXS (grazing incidence small angle X-ray scattering) in previous studies [28, 29]. For a random distribution atoms on an array of defects and if the clusters cannot escape from these defects, the size distribution is given by the Poisson law. The occurrence of a Poisson distribution for Pd clusters grown $Al_2O_3$ / $Ni_3Al(111)$ is supported by previous STM and GISAXS *in situ* characterization [28] and by Monte Carlo simulation [40]. The Poisson law gives for a deposit corresponding to a mean occupation of Q atoms per site (= number of deposited atoms divided by the number of sites) the probability P(Q, n) to have a nucleation center occupied by a cluster of n atoms:

$$P(Q, n) = (Q^n /n!) e^{-Q} \quad (1)$$

Table 1 gives P (Q, n) for deposits of 0.2 to 2 atoms per nucleation site. As seen in the introduction Pd clusters sit on the sites of the 'dot' structure. The density of nucleation sites is $n_s = 6.5 \times 10^{12}$ cm$^{-2}$. P (Q, 0) is the probability to have empty sites and the density of empty sites is $n_s \cdot P(Q, 0) = n_s \cdot e^{-Q}$, then the density of occupied sites is $n_s \cdot (1-e^{-Q})$. From table 1 we see that for the very small deposits we are concerned in this paper (0.2<Q< 2.5) not all sites are occupied.

| Q (atoms/site) | P(Q,0) | P(Q,1) | P(Q,2) | P(Q,3) | P(Q,4) | P(Q,5) |
|---|---|---|---|---|---|---|
| 0.2 | 0.8187 | 0.1637 | 0.0163 | 0.0011 | 5.46x10$^{-5}$ | 2.18x10$^{-6}$ |
| 0.5 | 0.6065 | 0.3032 | 0.0758 | 0.0126 | 0.0015 | 1.58x10$^{-4}$ |
| 1.0 | 0.3679 | 0.3679 | 0.1839 | 0.0613 | 0.0153 | 0.0031 |
| 1.5 | 0.2231 | 0.3347 | 0.2510 | 0.1255 | 0.0471 | 0.0141 |
| 2.0 | 0.1353 | 0.2707 | 0.2707 | 0.1804 | 0.0902 | 0.0361 |

Table 1: Probability P(Q,n) to have a cluster of size n for a deposit of Q atoms/site given by the Poisson distribution.

2.2 Reactivity measurements

The reactivity of Pd clusters has been studied at constant temperature (473 and 533K) with a background pressure of isotopically labelled oxygen ($^{18}O_2$) and a CO molecular beam. The CO beam, produced by supersonic expansion, is collimated in order that all the CO molecules impinge only the sample surface. The intensity of the CO beam ($J_{CO} = 2.0 \times 10^{13}$ molecules cm$^{-2}$ s$^{-1}$ and the equivalent pressure is $P_{co}= 6.8 \times 10^{-8}$ mbar) is modulated by a programmable shutter. The molecules coming out from the sample (CO and $CO_2$) are detected by a mass spectrometer. In order to have the highest sensitivity for the small amount of Pd on the sample (0.002 to 0.014 ML) the entrance of the mass spectrometer has been placed the closest as possible from the sample surface without occulting the CO beam. The mass spectrometer is working in a pulse counting mode. In order to have accurate and reproducible results we use rigorously the same experimental procedure which is illustrated by figure 1.

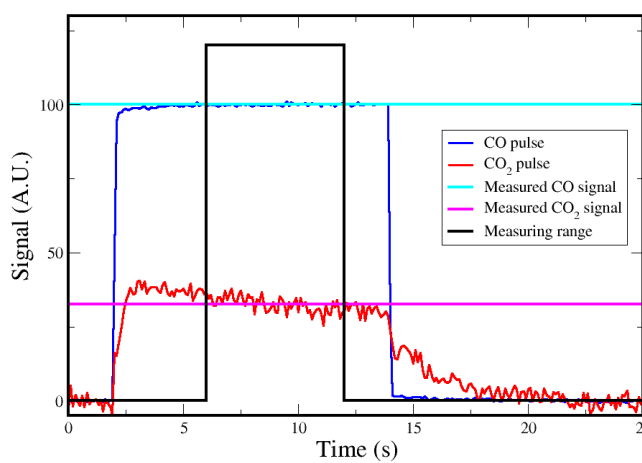

Figure 1: Example of experimental signals illustrating the experimental procedure for reactivity measurements. In blue: CO signal corresponding to CO desorbing from the sample before the introduction of $O_2$ (at T≥473 K the adsorption of CO on Pd clusters is fully reversible). In red: $CO_2$ signal (multiplied by a factor of 10) corresponding to the reaction, on the sample surface, of CO from the molecular beam with background $^{18}O_2$. In black: window where the steady state $CO_2$ production is obtained by the averaged value of the signal.

After deposition of Pd (at 373K) the sample is heated up to the set point temperature for the reaction (avoiding any overshoot that could induce mobility of the Pd). The intensity of the CO flux desorbing from the sample is first measured by averaging over 10 pulses of 12 seconds as illustrated in fig. 1. As at T≥473 K the adsorption of CO on Pd clusters is fully reversible, the desorbed flux is equal to the impinging flux. Then, a background pressure ($5 \times 10^{-8}$ mbar) of labelled oxygen ($^{18}O_2$) is introduced in the UHV chamber. To measure the reaction rate, a CO pulse of 12 s width in a cycle of 25 s is generated and the $^{12}C^{16}O^{18}O$ signal is continuously recorded. In order to increase the signal to noise ratio, the signal is averaged on 100 cycles for $CO_2$. The trigger is adjusted to start the acquisition of the signal two second before the opening of the CO beam in order to facilitate the subtraction of the $CO_2$ baseline (see figure 1). The steady state $CO_2$ production is obtained by adjusting by a constant the signal between 6 and 12 s (see figure 1). In order to correct from an evolution of the sensitivity of the mass spectrometer the $CO_2$ signal is divided by the intensity of the CO signal. The same experiment is repeated four times to check an eventual evolution of the $CO_2$ production. In fact (see figure S1, S2 in the supplementary material) at temperature larger than 493 K, the measured activity evolves between the first and the second experiment and then it is stable while at 473 K there is no noticeable change. We come back on this evolution in the discussion section. All this procedure is also carried out before the growth of Pd clusters in order to measure, in exactly the same conditions, the activity of the bare alumina film. This activity is weak but not zero and it has to be subtracted from the measured activity after Pd deposition. The intrinsic activity of the alumina film is 1.5 times smaller than the activity of a deposit of 1 atom/site at 473 K and 4.5 times smaller at 533K.

## III. Experimental results

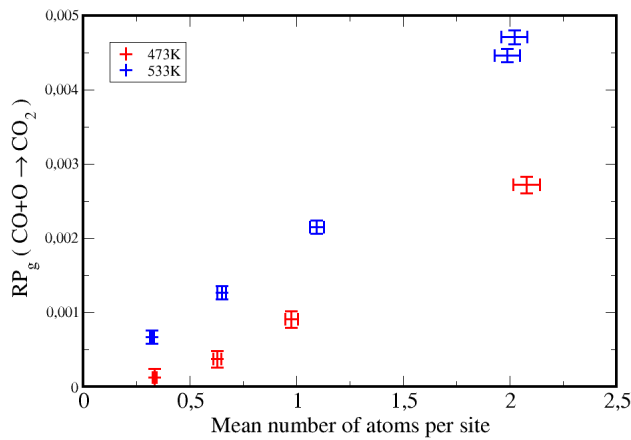

Figure 2: Activity in CO oxidation ($RP_g$) plotted as a function of the mean number of atoms per site (Q) at 473 K and 533 K

Figure 2 displays the activity in CO oxidation at T = 473 K and T = 533 K as a function of the mean number of atoms per site (Q). As the rate of $CO_2$ production is divided by the intensity of the CO beam ($J_{CO}$) impinging on the substrate this activity represents the probability that a CO molecule impinging on the sample reacts with adsorbed oxygen to form a $CO_2$ molecule. We call this term global reaction probability ($RP_g$). The normal reaction probability (RP) is the probability that one CO molecule striking the catalyst surface (Pd clusters in our case) forms a $CO_2$ molecule. RP is obtained by dividing the $RP_g$ by the Pd coverage ($\theta_{Pd}$) as long as the clusters stay two-dimensional (2D).

$$RP = RP_g / \theta_{Pd} \qquad (2)$$

The activity ($RP_g$) increases with temperature. One can remark that at 533K the activity seems to approach zero, more or less linearly when the Pd coverage decreases. It is not the case at 473 K where the reaction seems to require a minimum Pd coverage to start. For the small coverages used in these experiment, the clusters practically contain a maximum of 5 atoms (see table 1) then we can reasonably assume that all the atoms are exposed to the reacting gas and thus potentially can catalyze the CO oxidation.

In a general way, we can express the total activity A (number of $CO_2$ molecules produced per second and per unit surface of the sample) by the following equation:

$$A(Q) = n_s \cdot [a_1 P_1(Q) + a_2 P_2(Q) + a_3 P_3(Q) + a_4 P_4(Q) + a_5 P_5(Q) + .....] \quad (3)$$

In this equation $n_s$ is the density of sites which has been determined, by STM observations ($n_s = 6.5 \times 10^{12}$ cm$^{-2}$ for Pd on $Al_2O_3/Ni_3Al(111)$ [29]); $a_i$ and $P_i(Q)$ are the total activity of a cluster of i atoms (number of $CO_2$ molecules produced per second by a cluster of size i) and the probability to have a site occupied by a cluster of i atoms, respectively (then $n_s \cdot P_i$ is equal to the number of clusters of size i per unit surface of the sample). The quasi-linear evolution of the activity as a function of Q at 533K can be easily explained if the total activity of a cluster is proportional to the number of atoms it contains.

Indeed, if $a_i = i \cdot a$ equation 3 becomes:

$$A(Q) = n_s \cdot a \, [P_1(Q) + 2P_2(Q) + 3P_3(Q) + 4P_4(Q) + 5P_5(Q) + ......] = n_s \cdot a \cdot Q \quad (4)$$

A demonstration of the fact that a linear variation of A as function of Q implies that $a_i = i \cdot a$, is given in Supplementary Information. It is worthy to notice that this is valid for any size distribution of the clusters. The slope of the straight line (a) represents the activity per atom which is the TOF.

At 473 K, the evolution of the activity as a function of Q differs significantly from linearity. To understand this different behavior we make the following assumptions: (1) the deposition at 373K generates, as expected, a Poisson distribution; (2) some size classes of clusters do not contribute to the reaction; (3) for other clusters the activity is proportional to their number of atoms (what we learn from experiments performed at 533K).

If we consider that single atoms do not contribute to the reaction the activity will be:

$$A(Q) = n_s \cdot a \, [2P_2(Q) + 3P_3(Q) + 4P_4(Q) + 5P_5(Q) + ....] = n_s \cdot a \, (Q - P_1) \quad (5)$$

Now, considering that monomer and dimers do not contribute to the reaction the reactivity will be:

$$A(Q) = n_s \cdot a [Q - P_1(Q) - 2P_2(Q)] \quad (6)$$

Taking into account the first assumption of a Poisson distribution of clusters at 373K, $P_1(Q) = Q \cdot e^{-Q}$ and $P_2(Q) = \frac{1}{2} Q^2 \cdot e^{-Q}$.

In the experiment we measure the global reaction probability ($RP_g = A(Q)/J_{CO}$) instead of the activity A(Q). We have tried to fit the experimental data with equations 4, 5 and 6 which correspond to the case where all deposited atoms contribute to the reaction, only dimers and larger clusters are active and only trimers and larger clusters are active, respectively. We see on figure 3 that the data obtained at 533 K fit rather well with the first case where the activity per cluster is proportional to the number of atoms in the cluster and do not fit at all with the case where single atoms are not active nor with the case where single atoms and dimers are not active. On figure 4 we see that the data obtained at 473K behave differently, they fit only with the case where single atoms are not active for the reaction.

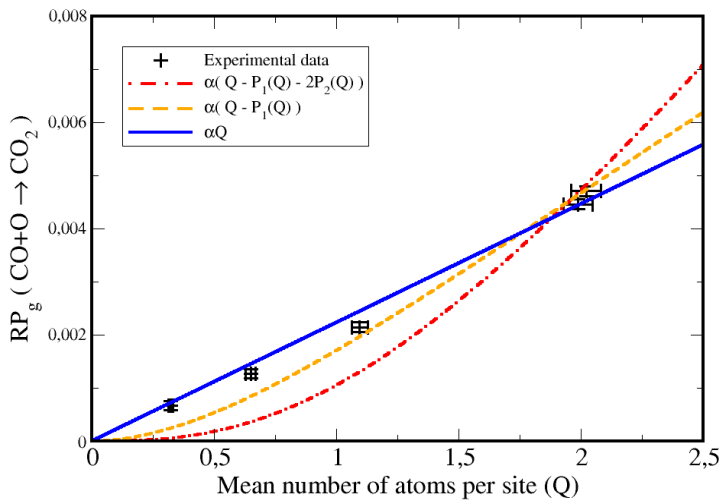

Figure 3: Fit of the activity in CO oxidation ($RP_g$) at 533 K, as a function of the mean number of deposited Pd atoms per sites (Q), with 3 functions: in blue (1) Q, in yellow (2) [Q - $P_1(Q)$], in red (3) [Q - $P_1(Q)$ – $2P_2(Q)$]. They correspond to three cases: (1) all atoms equally contribute to the reaction, (2) single atoms are not active for the reaction, (3) single atoms and dimers are not active. The best fit corresponds to the linear variation with α = 0.00223. $P_1(Q)$ and $P_2(Q)$ are given by the Poisson law. The fitting parameter α is equal to $a \cdot n_s /J_{CO}$.

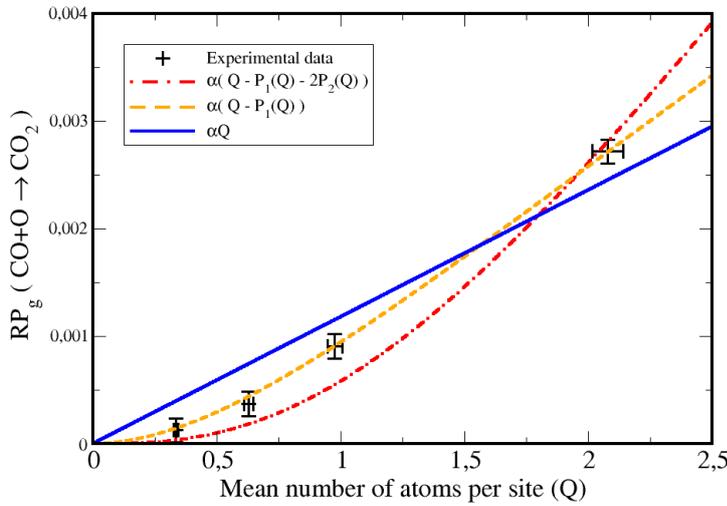

Figure 4: Fit of the activity in CO oxidation ($RP_g$) at 473 K, as a function of the mean number of deposited Pd atoms per sites (Q), with the three functions defined in fig.1. The best fit is obtained for the second function with α = 0.00149 which corresponds to the case where single atoms are not active. The fitting parameter α is equal to $a \cdot n_s /J_{CO}$.

From figure 3 the best fit (linear fit) is obtained for a slope α = 0.00223 which is proportional to a, the activity per atom, in eq. 4. From figure 4 the best fit is obtained with the model where the single atoms are not active, with α = 0.00149, meaning that only dimers and larger clusters contribute to the reaction.

The TOF is defined by the number of produced $CO_2$ molecules per second and per cm² divided by the number of Pd atoms per cm², it can be calculated by the following equation:

$$TOF = RP_g \cdot J_{CO}/Q \cdot n_s \qquad (7)$$

Then at T= 533 K, TOF = 3.077 α = 6.9x10⁻³ s⁻¹. At 473 K the TOF depends of Q it is equal to 4.6x10⁻³ x (1-e⁻ᵠ) s⁻¹.

Figure 5 represents the mean number of active atoms as a function of the mean number of atoms per site. Thus it confirms that at 473 K some atoms do not contribute to the reaction while at 533 K all deposited atoms do contribute. The difference between the blue and the yellow curve corresponds to the mean number of isolated atoms. We will discuss about this apparent contradictory results in the next section.

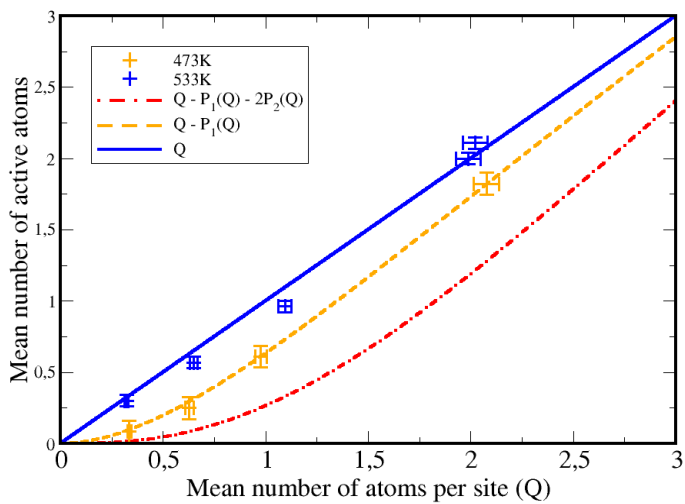

Figure 5: Mean number of active atoms in CO oxidation as a function of the mean number of atoms per sites (Q). The blue, yellow and red curves corresponds to the three cases defined in fig.3. The data points in blue and in yellow corresponds to T = 533 K and T = 473 K, respectively and they have been multiply by the reciprocal value of the α factors obtained from the best fits.

IV. Discussion

4.1. Stability of single atoms

The experimental results show that at 473 K only the $Pd_n$ clusters with $n \geq 2$ are active in CO oxidation while at 533 K all the deposited atoms contribute to the reaction. This apparent contradiction can be explained by assuming that at 533 K the originally isolated atoms are in a different configuration. The simplest interpretation is that at 533 K the single atoms are mobile and coalesce with other atoms or clusters up to the disappearance of all the initial single atoms. The mobility of single atoms was already suggested for $Pt_1$/alumina [6, 7] and $Pd_1$/alumina [10]. This mobility/coalescence explains the fact that at 533K the activity measured in the first run is smaller than in the other runs (we recall that we make successively four measurements in the same conditions) while at 473 K the same activity is obtained for the all successive measurements (see Supplementary Information). We can deduct from these observations that at 473 K the single atoms are stable while at 533 K single atoms are mobile and coalesce until only clusters ($Pd_n$, $n \geq 2$) are present. Let discuss now about the origin of the diffusion of Pd single atoms. Previous experiments have shown that $Pt_1$ is mobile on alumina under $CO+O_2$ or under $H_2$ and not in pure $O_2$ [6, 7]. Furthermore, the diffusion of Pt atoms on $TiO_2$ has been directly observed by STEM under $H_2$ and not under $O_2$ [8]. By the same technique the diffusion/coalescence of $Pt_1$ on $Fe_2O_3$ was observed at 523 K under pure CO and not under pure $O_2$ [9]. The diffusion of Pd single atoms, under CO, was also observed by STM on $Fe_3O_4$ (001) and it was established that the diffusing species was a Pd-carbonyl (PdCO) [18]. In the same experiment it was shown that Pd single atoms are stable under UHV up to 673K. STM experiments have also shown that CO induced the mobility of $Pt_1/Fe_3O_4$ [19]. From these previous experiments, it is most probable that in our experiments the mobility of Pd single atoms at 533 K was induced by the presence of CO.

4.2. Reaction mechanism

It has been observed that single atoms ($Pd_1$ or $Pt_1$) are active or not in CO oxidation when they are supported on different substrates. In our experiment we use $^{12}C^{16}O$ and $^{18}O_2$ and we detect $^{12}C^{16}O^{18}O$ which proves that oxygen molecules have been dissociated before reaction with CO. This observation corresponds to the Langmuir-Hinshelwood (LH) mechanism which has been observed, using molecular beam methods, for Pd (or Pt) extended surfaces [35] and for Pd clusters [31]. It is generally admitted that an $O_2$ molecule needs at least 2 contiguous metal atoms to be dissociated, then we can understand that the Langmuir-Hinshelwood mechanism is not possible on single atoms [3]. However, theoretical investigations have evidenced another possible mechanism for CO oxidation on Pd or Pt single atoms where one CO and one $O_2$ molecules are co-adsorbed on the catalytic atom and later form a $CO_3$ adsorbed species which further dissociates in an adsorbed O atom and a $CO_2$ molecule that readily desorbs [3, 13]. If this reaction would occur in our system on single Pd atoms we should have (at 473 K) a contribution of this reaction in the $^{12}C^{16}O^{18}O$ signal but it is not the case in our experiments because the fit with the function $[(Q - P_1(Q)]$ assuming that single atoms do not contribute to the $CO_2$ production would be impossible. Then we rule out this hypothesis. On reducible oxide supports ($TiO_2$, $Fe_2O_3$, $CeO_2$……) a CO oxidation mechanism different than the Langmuir-Hinshelwood one is observed [11, 36]. In this mechanism a CO molecule adsorbed on the catalytic metal reacts with an oxygen atom from the oxide support to form a $CO_2$ molecule, this mechanism is called Mars-van Krevelen (MvK). The MvK mechanism has never been observed on alumina which is not a reducible oxide. Nevertheless, we have checked the possibility of the MvK mechanism for Pd clusters on $Al_2O_3/Ni_3Al$ (111) (see Supplementary Information). For this purpose we synthesized the alumina film with $^{18}O_2$ instead of normal $^{16}O_2$ and we use $^{12}C^{16}O$ and $^{16}O_2$ for the reaction. We detected a $^{12}C^{16}O^{16}O$ coming from the LH mechanism and absolutely no $^{12}C^{16}O^{18}O$ coming from the MvK then we can exclude the possibility of the MvK mechanism in our system (see figure S3). The same conclusion was also drawn for CO oxidation on small Pt clusters on alumina ultrathin film on NiAl (110) [17].

4.3. Turnover frequency (TOF)

Figure 3 shows that at 533 K the catalytic activity varies almost linearly with the mean number of atoms per site (Q), then the TOF is proportional to the slope of the straight line (TOF = 3.077 ). The fact that the TOF is independent of the cluster size is understandable if we consider that our clusters ($Pd_n$, n = 2 to 5) are planar and each atom in the cluster has the same contribution. For 3D clusters, only surface atoms contributes to the reaction. Indeed, in a recent study Bourguignon and coworkers have shown, by sum frequency generation of CO adsorbed on $Pd_n$ clusters on $Al_2O_3/Ni_3Al$ (111), that the clusters are 2D for n ≤ 10 [37]. Another study by the group of Watanabe has later shown by STM that Pt clusters on $Al_2O_3$/NiAl (110) containing less than 20 atoms are 2D [17].

At 533K, the value of the TOF is 0.0069 $s^{-1}$, for $P_{O2}$ = $5 \times 10^{-8}$ mbar and $P_{CO}$ = $6.8 \times 10^{-8}$ mbar and the corresponding value of the reaction probability is RP = 0.34. Previously, we had studied the reactivity of Pd of clusters with a size of 181 ±13 atoms [30] and we had found, in the same conditions and without the reverse spillover correction [38], a TOF of 0.01 $s^{-1}$ which is larger ( but not so much) than the value obtained in the present study for much smaller clusters. We had shown [30] that at T ≥ 500K the TOF for Pd clusters of 181 atoms was the same than for a Pd (111) extended surface which was measured by Engel and Ertl in close experimental conditions [35]. Then we can conclude that $Pd_n$ clusters with 2 ≤ n ≤ 5 are less active than large clusters or extended surfaces. Anderson and coworkers [15] have measured on size selected Pd clusters (n = 5, 20, 25) deposited on alumina film, the steady state activity in CO oxidation, at low pressure ($P_{O2}$ = $1 \times 10^{-7}$ mbar and $P_{CO}$ = $6.6 \times 10^{-9}$ mbar). For $Pd_5$ they found at 533 K an activity per cluster of 0.02 leading to a TOF = 0.004 $s^{-1}$, a value which is close, but smaller, than those measured in the present study. In the case of Anderson et al. measurements the ratio $P_{O2}/P_{CO}$ (=15) is larger than in our case (0.74). However, both measurements are made in the O-rich regime of the reaction. In this regime the reactivity has a weak dependence in oxygen pressure [39] and a large positive dependence in CO pressure (the reaction is limited by the rapid desorption of CO) that could explain the smaller TOF in the Anderson et al. measurements [15]. Larger clusters (n = 20 or 25) are probably 3D then, without knowledge of the cluster shape it is difficult to deduce the TOF from the $CO_2$ production rate per cluster.

V. Conclusion

By precise measurements of the reactivity in CO oxidation of ultralow deposits ($2 \times 10^{-3}$ - $1.4 \times 10^{-2}$ ML) of Pd on a nanostructured alumina ultrathin film it has been possible to determine the steady state activity of small Pd clusters as a function of the number of atoms they contain. At 473 K single atoms are stable and not active in CO oxidation but at 533K they diffuse and coalesce forming larger clusters. The activity of $Pd_n$ clusters (n = 2-5) is proportional to the number of atoms they contain. The activity per atom, which corresponds to the TOF, is constant. At 533K the value of the TOF (0.007 $s^{-1}$) is close to those obtained for larger clusters containing 181± 13 atoms, and Pd (111) extended surface, in the same experimental conditions.